\documentclass[prl,aps,preprintnumbers]{revtex4}
\usepackage{epsfig,amssymb,amsfonts}

\usepackage{epsfig,amssymb,amsfonts}
\newcommand{\ket}[1]{\left | \, #1 \right\rangle}
\newcommand{\bra}[1]{\left \langle #1 \, \right |}

\input{Qcircuit}

\begin{document}

\title{Implementing universal quantum gates in coupled cavities}


\author{Dimitris G. Angelakis and Alastair Kay}{
  \address{Centre for Quantum Computation, Department
of Applied Mathematics
 and Theoretical Physics, University of Cambridge,
 Wilberforce Road, CB3 0WA, UK}
%
%

\begin{abstract}

We study a linear array of coupled cavities interacting with two
level systems and show how to construct stable, individually
addressable, qubits in this system from the long-lived atom-photon
excitations (polaritons) at each site. We derive the system dynamics
and show that is described by an XY Hamiltonian. We proceed by
showing how to implement non-local quantum gates and show that
combined with the inherent ability for individual addressing,
universal quantum computation is possible in this system. We finally
 discuss the prospects for experimental implementation
using various technologies involving dopants as  atoms, quantum dots
or Cooper pair boxes.
\end{abstract}

\maketitle


\subsection{Introduction}
A continuing problem in quantum information is to find physical
systems in which universal quantum computation can be implemented.
Existing proposals involve a wide variety physical systems including
linear optics, ion traps, NMR, quantum dots, superconductors,
neutral atoms in optical lattices, and flying atom schemes. Most of
these schemes are able to achieve a high degree of control over a
small number of systems, but are limited in their scalability. In
contrast to this, optical lattices have been extremely successful in
creating coherent states of a large number of atoms, but suffer from
the inability (or extreme difficulty) of performing operations on
individual atoms\cite{book1}.

On the other hand,
there have recently been theoretical and experimental breakthroughs
into the possibility of directly coupling high-Q cavities together, and in
achieving strong coupling between the cavity mode and an embedded
two-level system. A variety of technologies have been employed,
including fiber coupled micro-toroidal cavities interacting with atoms
\cite{aoki-2006,armani-vahala03}, arrays of defects in photonic band
gap materials (PBGs)
\cite{vuckovic-apl,song-noda05,AntonioBadolato05202005} and
superconducting qubits coupled through microwave stripline
resonators \cite{wallraff-2004}. This has prompted proposals for the
implementation of optical quantum computing
\cite{angelakis-ekert04}, the production of entangled photons
\cite{angelakis-bose06} and the realization of Mott insulating and
superfluid phases
\cite{angelakis-bose06b,hartmann-2006-2,greentree-2006}. Here we
consider the use of such arrays for the realization of universal
quantum computation.

\subsection{System} We start be describing the system and demonstrating how the complex energy level structure of hybrid light-matter excitations (polaritons) can be reduced to an effective two-level system. If we consider a chain of $N$ cavities (which is located in a defect) in 1D, then photons can propagate between the cavities, and this hopping mechanism induces a coupling between the cavities. A realization
of this has been studied in structures known as a coupled resonator
optical waveguides (CROW) or couple cavity  waveguides (CCW) in
photonic crystals, where it was shown that light propagation is
characterized by small dispersion, very small group velocities and
low losses. The dispersion relation can be derived in a simple way
by applying the tight binding formalism from the electronic band
theory. The eigenmode for the extended structure is given by
\begin{equation}
{\bf E}(r,k)=\sum_{n}e^{-ink\Lambda/N}E_{n}({ r})
\end{equation}
where $E_{n}( r)=E_{m}( r-(n-m)\Lambda)$ describe the localized ground state
modes for each cavity (Wannier functions) and $\Lambda$ is the distance
between the defects and the summation over $n$ includes all the
cavities. If we assume that $\Lambda=1$ for simplicity, then the dispersion relation is given by
\begin{equation}
\omega(k)=\omega_d[1+A \mbox{ cos} (2\pi k/N)],
\end{equation}
where $A$ is the tight binding parameter which depends on the
geometry and $\Lambda$, and $\omega_d$ is the frequency of an
individual defect.

The usual way to quantize a system is through its eigenmodes. The
Hamiltonian corresponding to the above dispersion relation is given
by

\begin{equation}
H=\omega_d \sum_{m} b^{\dagger}_{k}b_{k}[1+A \mbox{ cos }(2\pi k/N)]
\end{equation}
where $b^{\dagger}_{k}(b_{k})$ are the creation (annihilation) operators
of photons occupying the extended  eigenmode. We could equally well
describe the system dynamics using the operators of the localized
eigenmode (Wannier functions), $a^{\dagger}_{k}(a_{k})$. These
describe the creation (annihilation) of a photon in the localized
defect mode $k$.  It's straightforward to see that the
transformation connecting the two bases is given by
\begin{equation}
b_{k}=\frac{1}{N}\sum_{m=-N/2}^{N/2}a_{m}e^{-2\pi ikm/N}.
\end{equation}
In the localized mode basis, the Hamiltonian is given by
\begin{equation}
H=\sum_{k=1}^{N}\omega_d
a^{\dagger}_{k}a_{k}+\sum_{k=1}^{N}A(a^{\dagger}_{k}a_{k+1}+H.C.)
\end{equation}

Now assume that the cavities are doped with two level systems (atoms
or a quantum dots). We shall denote the two energy levels by $g_{N}$ and $e_{N}$ corresponding to the ground and
excited states of a dopant placed at defect $N$. The Hamiltonian
describing the system is
\begin{equation}
H=\sum_{k=1}^{N} H^{free}_{k}+H^{int}_{k}+H^{hop}_{k},
\end{equation}
where $H^{free}$ is the Hamiltonian for the free light and atom parts,
$H^{int}$ the Hamiltonian describing the internal coupling of the
photon and atom in a specific defect and $H^{hop}$ describes the light hopping between defects.
\begin{eqnarray}
H^{free}&&=\omega_{d}|1_{k}\rangle\langle 1_{k}|+\omega_{0}|e_{k}\rangle \langle e_{k}| \\
H^{int}&&=g (|1_{k}g_{k}\rangle\langle 0_{k} e_{k}|+H.C.)\\
H^{hop}&&= A(|1_{k}0_{k+1}\rangle \langle 0_{k} 1_{k+1}|
+|0_{k}1_{k+1}\rangle \langle 1_{k} 0_{k+1}|)
\end{eqnarray}
$\omega_{d}$ and $A$ are the photon frequencies and hopping rates
respectively and $g$ is the light-atom coupling strength. The
$H^{free}+H^{int}$ component of the Hamiltonian can be diagonalized
in a basis of combined photonic and atomic excitations, called {\it
polaritons}. These polaritons are defined by creation operators
$P_{k}^{(\pm,n)\dagger}=\ket{n\pm}_k\bra{g,0}_k$, where the
polaritons of the $k$th atom-cavity system are given by
$\ket{n\pm}_k=(\ket{g,n}_k\pm \ket{e,n-1}_k)/\sqrt2$ with energies
$E^{\pm}_{n}=n\omega_{d}\pm g\sqrt{n}$, and $\ket{n}_k$ denotes  the
$n$-photon Fock state. As has been shown elsewhere, a polaritonic
Mott phase exists in this system where a maximum of one excitation
per site is allowed \cite{angelakis-bose06b}. This originates from
the repulsion due to the photon blockade effect
\cite{PhysRevLett.79.1467,birnbaum-2005}. In this Mott phase, the
system's Hamiltonian can be written in the interaction picture as
\begin{eqnarray}
H_{I}=A\sum_{k=1}^{N-1}P^{\dagger}_{k}P_{k+1}+P_{k}P^{\dagger}_{k+1},
\end{eqnarray}
where $P_{k}^{\dagger}=P_{k}^{(-,1)\dagger}$. As double or more
occupancy of the sites is prohibited, one can identify
$P^{\dagger}_{k}$ with $\sigma^{+}_k=\sigma^x_k+i\sigma^y_k$, where
$\sigma^x_k$ and $\sigma^y_k$ are the standard Pauli operators. The
system's Hamiltonian then becomes the standard XY model of
interacting spin qubits with spin up/down corresponding to the
presence/absence of a polariton.
\begin{eqnarray}
H_I=A\sum_{k=1}^{N-1}\sigma^x_k\sigma^x_{k+1}+\sigma^y_k\sigma^y_{k+1}.
\end{eqnarray}
Some applications of XY spin chains in quantum information
processing can thus been implemented in this system \cite{bose}

\subsection{Implementing control phase gates}

Now that we have a system of effective two-level atoms, we can treat
these as qubits and try to find how to implement universal quantum
computation with them. By construction, single atoms can undergo
local rotations by the application of laser fields, and the lattice
sites can be spaced sufficiently far apart that lasers can
individually address each site. Thus, all we need to demonstrate is
an entangling gate between pairs of qubits. in order to achieve
this, we will use the natural systems dynamics of the XY Hamiltonian
which we have just demonstrated.

For a chain of 3 qubits, the Hamiltonian is
$$
H=A\left(\sigma_1^x\sigma_2^x+\sigma_1^y\sigma_2^y+\sigma_2^x\sigma_3^x+\sigma_2^y\sigma_3^y\right)
$$
and one can readily verify the following action on four of the states,
\begin{eqnarray}
e^{-iH\pi/(2\sqrt{2}A)}\ket{000}&=&\ket{000}    \nonumber\\
e^{-iH\pi/(2\sqrt{2}A)}\ket{001}&=&-\ket{100}   \nonumber\\
e^{-iH\pi/(2\sqrt{2}A)}\ket{100}&=&-\ket{001}   \nonumber\\
e^{-iH\pi/(2\sqrt{2}A)}\ket{101}&=&-\ket{101}.  \nonumber
\end{eqnarray}
This can be assigned the interpretation that if the central qubit is in the state $\ket{0}$, the state of the other two qubits is swapped, and both local $Z$-rotations and a controlled-phase gate are applied to them. Of central importance among these is the controlled-phase gate, precisely the entangling gate that we require. If the central qubit is in the $\ket{1}$ state, then the same effect is realised without the local $Z$-rotations \cite{yung-2004-4}. Thus, we do not in fact need to know the state of the central qubit, we simply need to measure it after the evolution time of $\pi/(2\sqrt{2}A)$ and by applying a measurement on the
`mediator' qubit (the middle of the chain of three) in the
$\sigma^z$ basis, a nonlocal gate results between the two extremal qubits
\cite{Christandl:2004a}. Depending on the measurement result,
$\ket{0}$ or $\ket{1}$, the operation performed between the two
computational qubits was either {\sc
SWAP}.$(\sigma^z\otimes\sigma^z).CP$ or {\sc SWAP}.$CP$
respectively. The local $\sigma^z$ rotations are readily compensated for, and the effect of the SWAP gates can be tracked to make sure the correct qubits interact when we want them to.
 This non-local gate combined with the
ability to perform individual rotation of the polaritonic qubits
allow for universal quantum computation in this system (the cavities
can be well separated in contrast to optical lattices
implementations for example). This can seen for example by
considering  the construction in Fig.~\ref{circuit} which allows,
with two applications of this gate, the construction of any
controlled-$U$ gate, where $U=A^\dagger
B^\dagger\sigma^zB\sigma^zA$. Hence we can separate the action of
the {\sc SWAP} and $CP$.

In order to implement this scheme experimentally, dissipation due to
spontaneous emission and cavity leakage need to be taken into
account.  As previously mentioned, there are three primary candidate
technologies; fiber coupled micro-toroidal cavities
\cite{aoki-2006,armani-vahala03}, arrays of defects in PBGs
\cite{vuckovic-apl,song-noda05, AntonioBadolato05202005} and
superconducting qubits coupled through microwave stripline
resonators \cite{wallraff-2004}. In order to achieve the required
limit of no more than one excitation per site
\cite{angelakis-bose06b}, the ratio between the internal atom-photon
coupling and the hopping of photons down the chain should be of the
order of $g/A\sim10^{2}-10^{1}$($A$ can be tuned while fabricating
the array by adjusting the distance between the cavities and g
depends on the type of the dopant). In addition, the cavity/atomic
frequencies to internal coupling ratio should be $\omega_d,\omega_0
\sim 10^{4}g,10^{5}g$ and the losses should also be small,
$g/\max(\kappa,\gamma)\sim 10^3$, where $\kappa$ and $\gamma$ are
cavity and atom/other qubit decay rates. The polaritonic states
under consideration are essentially unaffected by decay for a time
$10/A$ ($10$ns for the toroidal case and $100$ns for microwave
stripline resonators). The required parameter values are currently
on the verge of being realised in both toroidal microcavity systems
with atoms and stripline microwave resonators coupled to
superconducting qubits, but further progress is needed. Arrays of
defects in PBGs remain one or two orders of magnitude away, but
recent developments, and the integrability of these devices with
optoelectronics, make this technology very promising as well. In all
implementations the cavity systems are well separated by many times
the corresponding wavelength of any local field that needs to be
applied in the system for the measurement process.


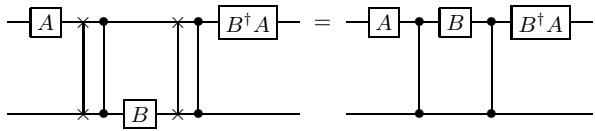
\begin{figure}
\resizebox*{0.45\textwidth}{!}{ $\begin{array}{ccc} \Qcircuit @C=1em
@!R {
& \gate{A} & \qswap & \ctrl{1} & \qw & \qswap & \ctrl{1} & \gate{B^\dagger A} & \qw \\
& \qw       & \qswap \qwx & \ctrl{-1} & \gate{B} & \qswap \qwx & \ctrl{-1} & \qw & \qw \\
}&=& \Qcircuit @C=1em @!R {
& \gate{A} & \ctrl{1} & \gate{B} & \ctrl{1} & \gate{B^\dagger A} & \qw  \\
& \qw   & \ctrl{-1} & \qw & \ctrl{-1} & \qw & \qw \\
}\end{array}$} \caption{The two-qubit gate we realise, in
conjunction with single-qubit rotations, is sufficient to create and
controlled-$U$ gate, and is hence universal.} \label{circuit}
\end{figure}

\subsection{Conclusions} In this paper, we have shown how universal quantum
computation could be realized in a coupled array of individually
addressable atom-cavity systems, where the qubits are given by mixed
light-matter excitations in each cavity site. While single-qubit
operations can be locally achieved, the only available interaction
between qubits is due to the natural system Hamiltonian. We show how
to manipulate this to give a controlled-phase gate between pairs of
qubits. This combined with the inherent ability of the system for
individual addressing allows for universal quantum computation. We
have discussed possible architectures for implementing these ideas
using photonic crystals, toroidal microcavities and superconducting
qubits and point out their feasibility and scalability with current
or near-future technology. \subsection{Acknowledgments}

 This work was supported by the
QIP IRC (GR/S821176/01), Clare College Cambridge and the European
Union through the Integrated
 Projects QAP (IST-3-015848), SCALA (CT-015714) and SECOQC.




\end{document}